\documentclass{ws-p10x7}
\def\lsim{\mathrel{\raise.3ex\hbox{$<$\kern-.75em\lower1ex\hbox{$\sim$}}}}
\def\gsim{\mathrel{\raise.3ex\hbox{$>$\kern-.75em\lower1ex\hbox{$\sim$}}}}
\newcommand{\nc}{\newcommand}
\nc{\prd}[3]{{\it Phys.\ Rev.}\ {{\bf D{#1}} #2 (#3)}}
\nc{\prl}[3]{{\it Phys.\ Rev.\ Lett.}\ {{\bf {#1}} #2 (#3)}}
\nc{\plb}[3]{{\it Phys.\ Lett.}\ {{\bf B{#1}} #2 (#3)}}
\nc{\npb}[3]{{\it Nucl.\ Phys.}\ {{\bf B{#1}} #2 (#3)}}
\nc{\ptp}[3]{{\it Prog.\ Theor.\ Phys.}\ {{\bf {#1}} #2 (#3)}}
\nc{\zpc}[3]{{\it Z.\ Phys.}\ {{\bf C{#1}} #2 (#3)}}
\nc{\mpla}[3]{{\it Mod.\ Phys.\ Lett.}\ {{\bf A{#1}} #2 (#3)}}
\nc{\rmp}[3]{{\it Rev.\ Mod.\ Phys.}\ {{\bf {#1}} #2 (#3)}}
\nc{\ijmpa}[3]{{\it Int.\ J.\ of\ Mod.\ Phys.}\
               {{\bf A{#1}} #2 (#3)}}
\nc{\app}[3]{{\it Acta\ Phys.\ Polon.}\ {{\bf B{#1}} #2 (#3)}}
\nc{\epj}[3]{{\it Eur. Phys. J.}\ {{\bf C{#1}} #2 (#3)}}
\nc{\prep}[3]{{\it Phys.\ Rep.}\ {{\bf {#1}} #2 (#3)}}
\begin{document}


\title{Search Strategies for non-Standard Higgses at $e^+e^-$ Colliders}

\author{J. Kalinowski}

\address{Instytut Fizyki Teoretycznej UW, ul.\ Ho\.za 69, 00681 Warsaw, 
Poland\\E-mail: kalino@fuw.edu.pl}

\twocolumn[\hfill IFT/00-23 
\maketitle\abstract{ The Higgs search strategies in
minimal non-supersymmetric extensions of the SM are discussed.}]

\section{Motivation}

If no new physics is assumed up to the grand unification or $M_{Pl}$
scales, the requirement of perturbativity and vacuum stability of the
Standard Model constraints\cite{per} the Higgs boson mass to lie
within the range of 130 -- 190 GeV. This is in perfect agreement with
the electroweak precision fits\cite{ew} which strongly point to a
light Higgs boson with $m_{H_{SM}}=62^{+53}_{-30}$ GeV, and with the
95\% CL upper limit 170 GeV.  This mass range, well above the ultimate
LEP2 reach (the current experimental LEP limit\cite{smhiggs} is
$m_{H_{SM}}>113.2$ GeV) and rather difficult at Tevatron (particularly
in its upper part), will be fully covered at the LHC by exploiting the
$gg\rightarrow H\rightarrow \gamma \gamma$ or associate production
$t\bar{t}H$, $WH$ processes.  For the future $e^+e^-$ colliders this
mass range is particularly easy.  The ``standard'' Higgs hunting
strategies at $e^+e^-$ collisions rely on the Higgs-strahlung,
$e^+e^-\rightarrow Z H$, and (for higher energies and heavier Higgs
bosons) on the $WW$ fusion, $e^+e^-\rightarrow \nu \bar{\nu}$,
processes \cite{lcrep}.

It should be stressed that the above implications for a light Higgs
boson {\it with} substantial $ZZh$ coupling can be altered if we admit
new physics. By adding ${\cal O}_i^{NEW}$ to the electroweak observables
${\cal O}_i$, the SM contributions can be compensated resulting in a
higher value of the Higgs mass.

In fact, the Higgs sector may turn out to be more complicated than
just one doublet, as realised in the SM.  Even in non-supersymmetric
world, and adding additional SU(2) singlet or doublet Higgs fields
only (to keep the tree-level $\rho=1$), Higgs boson couplings may
change considerably and thus complicate the Higgs boson searches.
Particularly worrisome is the case of a light Higgs $h$ with
suppressed $ZZh$ and $WWh$ couplings; we will refer to it as a
``bosophobic'' Higgs.  If such a Higgs boson with mass below 113 GeV
exists, negative searches at LEP2 in $e^+e^-\rightarrow ZH$ translate
into an upper limit on the $g_{ZZh}$ coupling.  Are we guaranteed to
discover the bosophobic Higgs with other Higgs bosons too heavy to be
produced? The answer turns out to be model dependent.  The absence of
$ZZ$ coupling implies that the $h$ will not be detectable at the
Tevatron, and very difficult, if not impossible, at the LHC. Therefore
we will consider a $\sqrt{s}=500-800$ GeV $e^+e^-$ linear collider
(LC) assuming an integrated luminosity $L\gsim 500$ fb$^{-1}$.

\section{Adding singlets}
Adding singlet Higgs fields does not pose any particular
theoretical problems nor benefits.  However, if many singlet fields
mix with the SM doublet in such a way that the physical Higgs bosons
$h_i$ share the SM WW/ZZ-Higgs coupling, the cross sections in
$e^+e^-\rightarrow Zh_i$ ($i=1,\dots,N$) for individual channels will
be suppressed.  The scenario considered in \cite{eg} assumes $h_i$
spaced more closely than the experimental mass resolution and spread
out over some substantial range around 200 GeV. 
The individual resonance peaks will
overlap making a diffuse signal not much different from the
background. If in addition Higgs bosons decay to a large number of
different channels, identification of individual final states will not
be possible nor useful due to large background.  Another possibility,
the so called stealthy Higgs, is considered in \cite{bb}, where the
usual Higgs doublet couples to many singlets (called Phions) which
interact among themselves strongly. The net effect is that the SM-like
Higgs boson is very broad and decays invisibly into Phions.

At hadron colliders such scenarios are real nightmare. On the other
hand, it has been demonstrated\cite{eg} that by looking for an excess
in the recoil mass $m_X$ distribution due to a ``continuum'' of
Higgses in $e^+e^-\rightarrow ZX$, the signal can be observed at an
$e^+e^-$ collider with $\sqrt{s}=500$ GeV and integrated luminosity
$>100$ fb$^{-1}$.  Since the inclusive $e^+e^-\rightarrow ZX$ process
can be used irrespectively of Higgs decay modes, the stealthy Higgs
can cleanly be detected\cite{bb} by looking for a signal of leptons
and missing energy.

\section{Adding one Higgs doublet}
Even the simplest two-Higgs-doublet model (2HDM) extension of the SM
exhibits a rich Higgs sector structure.  The CP-conserving (CPC) 2HDM
predicts the existence of two neutral CP-even Higgs bosons ($h^0$ and
$H^0$, with $m_{h^0}\leq m_{H^0}$ by convention), one neutral CP-odd
Higgs ($A^0$) and a charged Higgs pair ($H^\pm$). The same spectrum of
Higgs bosons is found in the minimal supersymmetric model (MSSM),
where it has been demonstrated\cite{gsw} that the detection of at
least one of the Higgs bosons is possible either at LEP2 or LHC.

The situation is more complex in the non-supersymmetric 2HDM.  Here we
consider the type-II 2HDM, wherein one of the doublets couples to
down-type quarks and leptons and the other to up-type quarks. The 2HDM
allows for spontaneous and/or explicit CP violation (CPV) in the
scalar sector\cite{lw} at the tree level.  In the CPV case the
physical mass eigenstates, $h_i$ ($i=1,2,3$), are mixtures (specified
by three mixing angles $\alpha_i$, in addition to the mixing angle
$\beta$ related to Higgs vev's) of the real and imaginary components
of the original neutral Higgs doublet fields; as a result, the $h_i$
have undefined CP properties.

\begin{figure}
\epsfxsize150pt
\figurebox{200pt}{300pt}{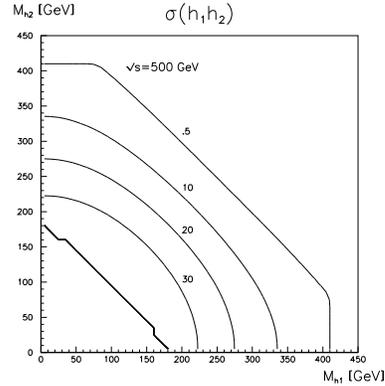}
\caption{Contour lines for ${\rm
min}[\sigma(e^+e^-\rightarrow h_1 h_2)]$ in units of fb's.  The
contour lines are plotted for $\tan\beta=0.5$; the plots are virtually
unchanged for larger values of $\tan\beta$. The contour lines overlap in
the inner corner as a result of excluding mass choices inconsistent
with experimental constraints from LEP2 data. 
[From \protect\cite{ggk1}]}
\end{figure}

If there are two light Higgs bosons $h_1$ and $h_2$, in the sense that
$Zh_1$, $Zh_2$ and $h_1h_2$ channels are kinematically open, then at
least one will be observable in $Zh_1$ or $Zh_2$ production or both in
$h_1h_2$ pair production. This is because of the sum rule\cite{gbsum}
for the couplings of any two of neutral Higgses to the $Z$ boson 
\begin{equation}
C_i^2+C_j^2+C_{ij}^2=1,
\label{oldsr} 
\end{equation}
where $g_{ZZh_i} \equiv \frac{g m_Z}{c_W} C_i$ and $g_{Zh_ih_j} \equiv
\frac{g}{2c_W} C_{ij}$, which says that all three couplings cannot be
simultaneously suppressed.  For example, if both $C_1$ and $C_2$ are
dynamically suppressed, then from the above sum rule it follows that
Higgs pair production is at full strength, $C_{12}\sim 1$. In Fig.1
contour lines are shown for the minimum value of the pair production
cross section, $\sigma(e^+e^- \rightarrow h_1 h_2)$ as a function of
Higgs boson masses.  The mimimum of $\sigma(h_1h_2)$ is
found\cite{ggk1} by scanning over the mixing angles $\alpha_i$
consistent with present experimental constraints on $C_i$ (which
roughly exclude $m_{h_1}+m_{h_2}\lsim 180$ GeV) and the assumption of
less that 50 $Zh_i$ events.  With $L=500$ fb$^{-1}$ a large number of
events is predicted for a broad range of Higgs boson masses.  If 50
$h_1h_2$ events before cuts and efficiencies prove adequate
(i.e. $\sigma>0.1$ fb), one can probe reasonably close to the
kinematic boundary.

The main question, however, is whether a {\it single} neutral Higgs
boson $h_1$ will be observed in $e^+e^-$ collisions if it is
sufficiently light, regardless of the masses and couplings of the
other Higgs bosons.  Such a scenario can easily be arranged by
choosing model parameters so that the $ZZ/WWh_1$ couplings are too
weak for its detection in Higgs-strahlung or WW fusion processes, and
all other Higgs bosons are too heavy to be produced via $Zh_i$ or
$h_ih_j$ processes at a given energy. In the CPC for example, one can
simply choose $h_1=A^0$ (the tree-level $ZZ/WWA^0$ coupling is zero),
or in the general CPV model choose mixing angles $\alpha_i$ to zero
the $ZZ/WWh_1$ coupling.  Since the other Higgs bosons are assumed to
be quite heavy to avoid production, implying no {\it light} Higgs with
substantial $ZZ/WW$ couplings, it would seem that the fit to precision
electroweak constraints is likely to be poor.  However, as shown in
\cite{ckz}, a good global fit to EW data is possible even for very
light $h^0$ or $A^0$ (with $m\sim 20$ GeV) in the CPC 2HDM.

If one of the two processes, $Zh_2$ and $h_1h_2$, is beyond the LC's
kinematical reach, the sum rule in Eq.~(\ref{oldsr}) is not sufficient
to guarantee $h_1$ discovery if $C_1\ll1$.  However, in this case we
can exploit other sum rules\cite{ggk1} which constrain the Yukawa
couplings of any Higgs boson $h_i$.  For $C_i\ll1$ they read (for
obvious reasons we consider the third generation fermions)
\begin{eqnarray}
  \label{yuksr}
({S}^t_i)^2 + ({P}^t_i)^2 
&=&\cot^2\beta 
\nonumber \\
({S}^b_i)^2 + ({P}^b_i)^2 
&=&\tan^2\beta 
\end{eqnarray}
where the fermionic Higgs couplings are given by
$\frac{gm_f}{2m_W}\bar{f}(S^f_i +i\gamma_5P^f_i)fh_i$, i.e. $S^f_i$
and $P^f_i$ are defined relative to the SM strength.  Combining the
two sum rules we find that the Yukawa couplings to top and bottom
quarks cannot be simultaneously suppressed, {\it i.e.}  at least one
$h_i$ Yukawa coupling must be large. Therefore the Higgs hunting
strategies should include not only the Higgs-strahlung and Higgs-pair
production but also Yukawa processes with Higgs radiation off top and
bottom quarks in the final state.  The current experimental limits in
the $m_{h_1}$-$\tan\beta$ parameter space are rather weak, see
\cite{kzm}.

It turns out that for large (small) $\tan\beta$, the $b\bar{b}h_1$
($t\bar{t}h_1$) cross sections are comfortably large for $h_1$
discovery. However, scanning over mixing angles $\alpha_i$ we
find\cite{ggk2} the difficult region of moderate $\tan\beta$, where
even at very high integrated luminosity of 2500 fb$^{-1}$ none of the
$Zh_1$, $t\bar{t}h_1$ and $b\bar{b}h_1$ processes yields more than 50
events, see Fig.2.

\begin{figure}
\epsfxsize150pt
\figurebox{200pt}{300pt}{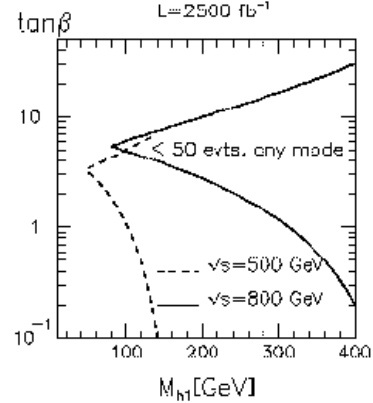}
\caption{ The maximum and minimum $\tan\beta$ values between which
$t\bar t h_1$, $b\bar b h_1$ and $Zh_1$ final states all have fewer
than 50 events assuming $L=2500$ fb$^{-1}$ at $\protect\sqrt{s}=500$
GeV (dashes) and $\protect\sqrt{s}=800$ GeV (solid).  Masses of the
remaining Higgs bosons are assumed to be 1000 GeV. [From
\protect\cite{ggk2}]}
\end{figure}

The non-discovery wedge begins at $m_{h_1}\sim 50 $ GeV at
$\sqrt{s}=500$ GeV ($\sim 80$ GeV for $\sqrt{s}=800$ GeV) and expands
rapidly as $m_{h_1}$ increases.  Thus, it is apparent that, despite
the sum rules guaranteeing significant fermionic couplings for a light
2HDM Higgs boson that is unobservable in $Z$+Higgs production,
$\tan\beta$ and the $\alpha_i$ mixing angles can be chosen so that the
cross section magnitudes of the two Yukawa processes are
simultaneously so small that detection of such an $h_1$ cannot be
guaranteed for integrated luminosities that are expected to be
available.

Is the whole wedge consistent with electroweak constraints? This
question, in the context of CPC 2HDM, is analysed in \cite{cfggkk}
with the general result that for LC $\sqrt{s}=500$ (800) GeV, the
$\tan\beta\sim 2$ portions of the 2HDM no-discovery wedges in
$m_{h_1}$-$\tan\beta$ parameter space have $\Delta\chi^2<1$ ($<1.5$)
(relative to the best SM fit) and all of the no-discovery wedges'
portions with $\tan\beta\gsim 1$ have $\Delta\chi^2<2$.
Thus the discrimination from current EW data
between the SM and the no-discovery scenarios in the 2HDM
is rather weak at the LC with $\sqrt{s}=500-800$ GeV.

\section{Conclusions}
In a general CPV 2HDM a light bosophobic Higgs boson, with all other
Higgs bosons heavier than the kinematical reach of a 500-800 GeV
$e^+e^-$ collider, may escape detection. If $\sqrt{s}$ is pushed
beyond 1 TeV, and the next lightest Higgs $H$ is still not seen in
$ZH$ or $\nu\bar{\nu}H$, implying $m_H\gsim 1$ TeV, one would expect
to see strong $WW$ scattering behavior at both the LHC and the LC.  As
a result, only an LC with sufficiently large energy to probe a
strongly interacting $WW$ sector could be certain of seeing a Higgs
signal, unless the electroweak fits really {\it do} indicate a
relatively light Higgs boson.

\section*{Acknowledgments}
I am grateful to P.~Chankowski, B.~Grzadkowski, J.~Gunion, M.~Krawczyk 
and P.~Zerwas for many discussions. Work partially supported by the KBN 
Grant  2 P03B 052 16.

\end{document}